\documentclass{PoS}
\usepackage{amsmath}
\usepackage{color}

\def\muf{{\mu^{}_F}}

\def\mur{{\mu^{}_R}}

\def\schannel{\mbox{$s$-channel}}
\def\ttbar{$t\bar t$}

\newcommand{\Hathor}{\textsc{HatHor}}
\newcommand{\msbar}{$\overline{\mathrm{MS}}\, $}

\title{
\vspace*{-3.3cm}
\begin{minipage}{\textwidth}
{\normalfont\small DESY 16-118
\hspace{\fill} July 2016}\\
\end{minipage}\\[60pt]
Single-top production in the \schannel\ and the top-quark mass}

\ShortTitle{Single-top production in the \schannel\ and the top-quark mass}

\author{Sergey Alekhin$^{\,\,ab}$, Sven-Olaf Moch$^{\,\,a}$, and \speaker{Stephan Thier}$^{\,\,a}$\\
        \llap{$^a$}II. Institut f{\"u}r Theoretische Physik,
        Universit{\"a}t Hamburg, Luruper Chaussee 149, \\ D-22761 
        Hamburg, Germany\\
        \llap{$^b$}Institute for High Energy Physics, 142281 Protvino,
        Moscow region, Russia\\
        E-mail: \email{sergey.alekhin@desy.de}, \email{sven-olaf.moch@desy.de},
        \email{stephan.christoph.thier@desy.de}}

\abstract{We use a fixed-order expansion of resummed soft-gluon corrections
  to determine an approximate NNLO formula for the
  partonic cross section of single-top production
  in the \schannel. This formula is implemented in
  the program Hathor for the numerical evaluation
  of hadronic cross sections. With the resulting code,
  we perform a fit of the top-quark mass to
  Tevatron cross section data.
  Results for $m_t$ are given in the pole-mass scheme
  and in the \msbar\ scheme.}

\FullConference{XXIV International Workshop on Deep-Inelastic Scattering and Related Subjects\\
                 11-15 April, 2016\\
                 DESY Hamburg, Germany}

\begin{document}

\section{Introduction}

Single-top production proceeds via an electroweak interaction,
which puts it into contrast to top-quark pair production
where strong interactions generate the largest part
of top-quarks that are produced in hadron collisions.
Due to the different interactions, single-top production
can be used as a complementary way to access
top-quark parameters with minimal dependence
on strong interactions.
This complementary perspective can be enhanced
by considering the \schannel\ of single-top production,
which proceeds at leading order (LO) via 
quark--anti-quark annihilation to a $W$ boson that subsequently
decays into a top-quark and, in most cases, a bottom quark.
There is no contribution of initial-state gluons to
this process in contrast to their dominant contribution
to top-quark pair production.

While fixed-order corrections for single-top production in the \schannel\
are known to next-to-leading order (NLO) in perturbative QCD~\cite{Smith:1996ij},
further soft-gluon corrections have been
considered and were found to be
sizeable~\cite{Kidonakis:2006bu,Kidonakis:2007ej,Kidonakis:2010tc}.
The emission of soft gluons generates
logarithmic terms which become large near the production threshold.
After resummation of these logarithms,
an expansion in powers of the strong coupling
allows to extract important QCD corrections
at fixed orders in perturbation theory beyond NLO.

In this paper, we calculate an approximate next-to-next-to-leading order (aNNLO)
formula for the partonic cross section of single-top production
in the \schannel\ based on the fixed-order expansion of
soft-gluon corrections which was presented in~\cite{Kidonakis:2010tc}.
Subsequently, we implement this formula in
the program \Hathor~\cite{Aliev:2010zk,Kant:2014oha}
to evaluate the hadronic cross section numerically.
With this code, we perform fits to Tevatron cross-section data
and determine the top-quark mass
in both the pole-mass scheme and the \msbar\ scheme.

\section{Calculation}

We consider the partonic cross section $\sigma$ in perturbation theory as a
power series in the strong coupling $\alpha_s$,
\begin{equation}
\label{eq:sigma}
\sigma \,=\, \sigma^{(0)} + \sigma^{(1)} + \sigma^{(2)}\, ,
\end{equation}
with the LO partonic cross section for the process 
$u\bar{d} \to t\bar{b}$ given by 
\begin{equation}
\label{eq:sigma0}
\sigma^{(0)} =
\frac{\pi  \alpha^2 V_{tb}^2 V_{ud}^2 (m_t^2-s)^2 (m_t^2+2 s)}
{24 s^2 \sin^4\theta_W (m_W^2-s)^2}
\end{equation}
and the exact NLO result $\sigma^{(1)}$ computed in~\cite{Smith:1996ij},
while $\sigma^{(2)}$ in Eq.~(\ref{eq:sigma}) is currently unknown.
Here $s$ is the partonic center-of-mass energy squared, 
$m_t$ and $m_W$ are the top-quark and $W$-boson masses respectively,
and $\alpha$, $\sin\theta_W$, $V_{tb}$ and $V_{ud}$
are the electro-weak and CKM parameters~\cite{Kant:2014oha}.

To validate our approach, we start with the
approximate next-to-leading order (aNLO)
double-differential cross section of Ref.~\cite{Kidonakis:2006bu}
for single-top production in the \schannel\
and compare the results of our procedure
to the complete NLO corrections that are implemented in \Hathor.
First, we perform an analytic integration
over the Mandelstam variables $t$ and $u$ 
to obtain the partonic cross section as a function 
of the top-quark velocity $\beta = (1 - m_t^2/s)^{1/2}$ only.
In accordance with the soft-gluon approximation,
that is valid near the production threshold,
we keep only the lowest order in $\beta$
during this integration.
To avoid overestimating contributions at large $\beta$ in this approach, we multiply the result 
by a kinematical suppression factor $1-\beta^2 = m_t^2/s$
and find 
$\sigma^{(1)} \simeq R_{\text{aNLO}} \sigma^{(0)}$
with
\begin{equation}
\label{eq:RaNLO}
R_{\text{aNLO}} =
\frac{\alpha_s C_F}{8 \pi } \left(1-\beta^2\right) \left(
  112 \log ^2(\beta )
  -148 \log (\beta )
  +63
  -4 \log\left(\frac{\mu_F^2}{m_t^2}\right) (8 \log (\beta )-3) 
\right)
+\mathcal{O}(\beta)\, ,
\end{equation}
where the strong coupling $\alpha_s=\alpha_s(\mu_R)$ is taken at the
renormalization scale $\mu_R$, which is kept separate from the 
factorization scale $\mu_F$.

For numerical evaluation of hadronic cross sections,
we implement our result in the program \Hathor, that was developed
for the evaluation of the inclusive cross section in both
\ttbar\ production~\cite{Aliev:2010zk}
and single-top production~\cite{Kant:2014oha}.
To parametrize the partonic content of nucleons,
the ABM12 parton distribution functions~\cite{Alekhin:2013nda} are employed.

\begin{figure}[t]
\begin{center}
\includegraphics[width=0.65\textwidth]{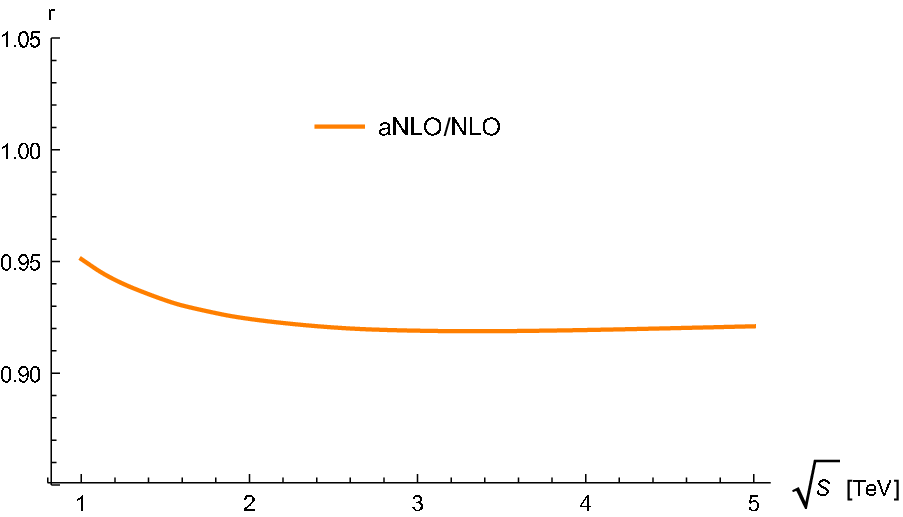}
\caption{Ratio of the aNLO formula to the exact NLO
result $r = (R_{\text{aNLO}} \sigma^{(0)})/\sigma^{(1)}$ for the cross section of single-top production 
in the \schannel\ as function of the hadronic center-of-mass energy.
}
\label{figure_aNLO_NLO_comparison}
\end{center}
\end{figure}

The ratio of the cross section in our aNLO approximation
to the complete NLO result for different
hadronic center-of-mass energies is given in Fig.~\ref{figure_aNLO_NLO_comparison}.
All results presented here refer to $p\bar{p}$ collisions
with a pole mass $m_{t,\text{pole}} = 172.5$~GeV.
While the agreement of our approximation with the exact result
is quite good over the considered energy range,
it gets better at smaller energies
as expected for the threshold approximation that forms
the foundation of our calculation.
This validation justifies application of the same procedure
at NNLO to obtain an estimate for the
corrections at the next order in perturbation theory.

To determine the partonic cross section at aNNLO,
we follow the same steps that were
described above at aNLO using the approximate NNLO
double-differential partonic cross section of Ref.~\cite{Kidonakis:2010tc}, 
which is accurate to the next-to-next-to-leading logarithm.
In this way, we obtain exact expressions for 
the leading terms, $\log^4(\beta)$ and $\log^3(\beta)$, 
and terms proportional to $\log^2(\beta)$ except for the interference with the 
term $\sim\beta^0$ at NLO.
We find $\sigma^{(2)} \simeq R_{\text{aNNLO}} \sigma^{(0)}$ with
\begin{multline}
\label{eq:RaNNLO}
R_{\text{aNNLO}}= 
\frac{\alpha_s^2 C_F}{24 \pi ^2} \left(1-\beta^2\right) \Bigg(
  2352 C_F \log^4(\beta )
  -8 \log ^3(\beta ) (17 \beta_0+777 C_F)
\\
  +\frac{1}{3} \log ^2(\beta ) \left(801 \beta_0 -28 \left(3 \pi ^2-67\right)
  C_A +24759 C_F -504 \pi ^2 C_F -280 n_f+\frac{144}{N_c}\right)
\\
  +\log \left(\frac{\mu_F^2}{m_t^2}\right) \bigg(
    -1344 C_F \log ^3(\beta )
    +12 \log ^2(\beta ) (7 \beta_0+190 C_F)
\\
    -\frac{1}{3} \log (\beta )
    \left(333 \beta_0-8 \left(3 \pi^2-67\right) C_A+6066 C_F-144\pi^2 C_F-80 n_f\right)
\\
    +\log\left(\frac{\mu_R^2}{\mu_F^2}\right) (-24 \beta_0 \log (\beta )+18 \beta_0)
  \bigg)
\\
  +\log ^2\left(\frac{\mu_F^2}{m_t^2}\right) \left(
    192 C_F \log ^2(\beta )
    -12 \log(\beta )(\beta_0+12 C_F)
    +3 (3\beta_0+20 C_F)
  \right)
\\
  +\log\left(\frac{\mu_R^2}{\mu_F^2}\right) \left(
    84 \beta_0 \log ^2(\beta )-111\beta_0 \log (\beta )+\frac{189}{4} \beta_0
  \right)
\Bigg)
+\mathcal{O}(\beta), \hspace{25mm}
\end{multline}
where $\beta_0 = (11 C_A - 2 n_f)/3$,
$n_f$ is the number of quark flavors,
$C_F=\frac{4}{3}$, $C_A=3$, and
$N_c=3$ is the number of colors in QCD.

\begin{figure}
\begin{center}
\includegraphics[width=0.65\textwidth]{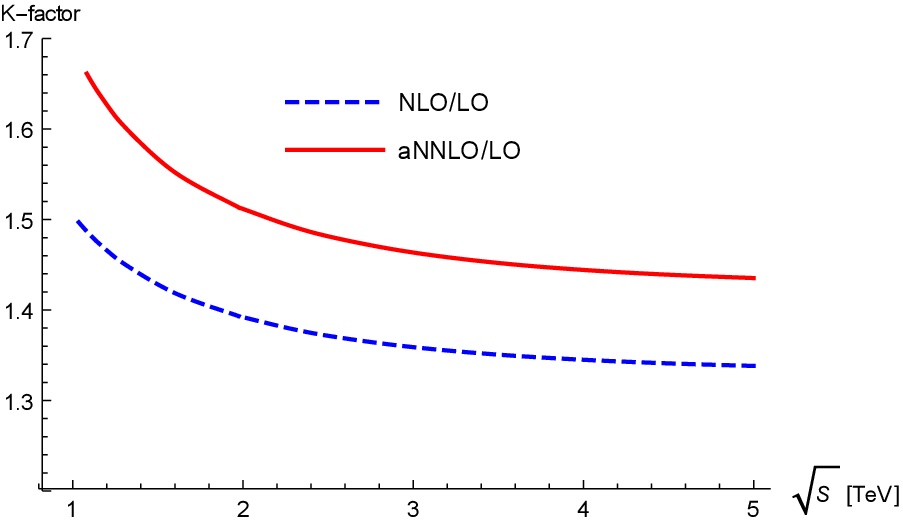}
\caption{The $K$-factors of the cross section for single-top production in the \schannel\
relative to the cross section at LO as function of the hadronic center-of-mass energy
at NLO $(\sigma^{(0)} + \sigma^{(1)})/\sigma^{(0)}$ (blue, dashed) 
and aNNLO $(\sigma^{(0)} + \sigma^{(1)} +  R_{\text{aNNLO}} \sigma^{(0)})/\sigma^{(0)}$ (red, solid).
}
\label{figure_aNNLO_NLO_comparison}
\end{center}
\end{figure}

Hadronic cross sections at different energies
from the implementation of our aNNLO result in \Hathor\
are given in Fig.~\ref{figure_aNNLO_NLO_comparison}
for the nominal scale choice $\mur = \muf = m_{t,\text{pole}}$ 
and compared to the exact NLO result.
In contrast to the large NLO corrections,
the approximate NNLO corrections are moderate
thus demonstrating apparent convergence 
of the perturbative expansion.
At the Tevatron center-of-mass energy \mbox{$\sqrt{S} = 1.96$~TeV},
the cross section is enhanced by $+39\%$ at NLO
compared to LO. When our aNNLO corrections are included,
this enhancement grows to $+51\%$ at the nominal scale.

\begin{figure}
\begin{center}
\includegraphics[width=0.65\textwidth]{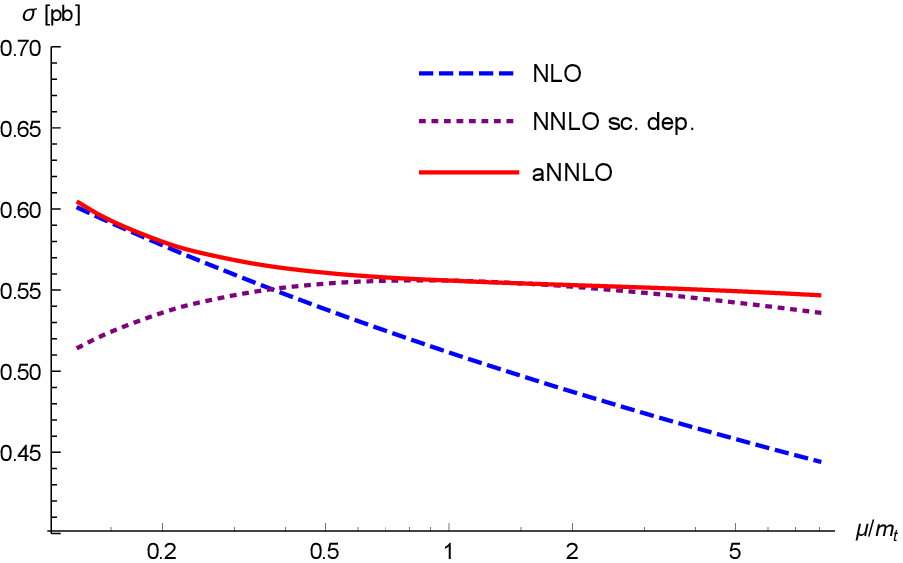}
\caption{Cross section of single-top production in the \schannel\ at $\sqrt{S} = 1.96$~TeV
as function of $\mu/m_t$ with $\mu = \mur = \muf$
at NLO (blue, dashed), aNNLO (red, solid),
and aNNLO with scale dependence exact at NNLO (purple, dotted).
}
\label{figure_aNNLO_NLO_NNLO_scale_dependence}
\end{center}
\end{figure}

As a cross-check of the perturbative stability of our aNNLO results,
we also study their dependence on unphysical scales.
In Fig.~\ref{figure_aNNLO_NLO_NNLO_scale_dependence},
we compare the scale dependence at NLO 
to the one at aNNLO based on Eq.~(\ref{eq:RaNNLO}) 
and to the exact NNLO scale dependence
that is included in \Hathor.
Both the renormalization scale $\mur$
and the factorization scale $\muf$
are varied simultaneously by a factor
between $1/8$ and $8$.
When both scales are varied independently,
it becomes apparent that our approximation
does not include all scale-dependent terms.
In this case, variations of uncancelled terms
lead to a scale dependence
which is similar in magnitude to the one at NLO.

\section{Mass Fit}

We use the implementation of our aNNLO formula for
\schannel\ single-top production in \Hathor\ to
extract the top-quark mass from the total cross section
which was measured at the Tevatron.
Combined observations by CDF and D0~\cite{CDF:2014uma}
allowed to determine $\sigma_s = 1.29^{+0.26}_{-0.24}$~pb
with a statistical significance of 6.3 standard deviations.

Our result for the top-quark pole mass
at aNNLO accuracy is
$m_{t,\text{pole}} = 166.8^{+8.0}_{-6.9}$~GeV.
Since the top-quark is a color-charged particle,
strong dynamics limit the accuracy with which
a pole mass can be defined.
It is thus advantageous to consider other
mass definitions which follow a theoretically
well-defined prescription, e.g.\ the \msbar\ mass.
We implement in our code a translation from
the pole mass to the \msbar\ mass at NNLO accuracy~\cite{Melnikov:2000qh}
and perform a mass fit which leads to
$m_{t,\overline{\mathrm{MS}}} = 159.2^{+7.5}_{-6.6}$~GeV
at the scale $\mu_R=m_{t,\overline{\mathrm{MS}}}$.

\section{Conclusions}

We have used a fixed-order expansion of resummed soft-gluon corrections to
\schannel\ single-top production in order to derive 
a compact analytic formula which approximates the
partonic cross section at NNLO.
Based on the implementation of these results in \Hathor,
hadronic cross sections have been evaluated
and the impact of the approximate NNLO corrections
was found to be moderate.
Fits of the top-quark mass in the pole-mass scheme
and in the \msbar\ scheme to the total cross section
exhibit a tendency to smaller mass values compared to
conventional methods for the extraction of $m_t$
however the results are compatible within uncertainties.

\acknowledgments
This work has been supported by Deutsche Forschungsgemeinschaft in
Sonderforschungs\-be\-reich SFB 676.

\bibliography{proceedings_single_top_bib}

\providecommand{\href}[2]{#2}\begingroup\raggedright\begin{thebibliography}{1}

\bibitem{Smith:1996ij}
M.~C. Smith and S.~Willenbrock, \emph{{QCD and Yukawa corrections to single top
  quark production via $q \bar{q} \to t \bar{b}$}},
  \href{http://dx.doi.org/10.1103/PhysRevD.54.6696}{\emph{Phys. Rev.} {\bf D54}
  (1996) 6696--6702}, [\href{http://arxiv.org/abs/hep-ph/9604223}{{\tt
  hep-ph/9604223}}].

\bibitem{Kidonakis:2006bu}
N.~Kidonakis, \emph{{Single top production at the Tevatron: Threshold
  resummation and finite-order soft gluon corrections}},
  \href{http://dx.doi.org/10.1103/PhysRevD.74.114012}{\emph{Phys. Rev.} {\bf
  D74} (2006) 114012}, [\href{http://arxiv.org/abs/hep-ph/0609287}{{\tt
  hep-ph/0609287}}].

\bibitem{Kidonakis:2007ej}
N.~Kidonakis, \emph{{Higher-order soft gluon corrections in single top quark
  production at the LHC}},
  \href{http://dx.doi.org/10.1103/PhysRevD.75.071501}{\emph{Phys. Rev.} {\bf
  D75} (2007) 071501}, [\href{http://arxiv.org/abs/hep-ph/0701080}{{\tt
  hep-ph/0701080}}].

\bibitem{Kidonakis:2010tc}
N.~Kidonakis, \emph{{NNLL resummation for s-channel single top quark
  production}}, \href{http://dx.doi.org/10.1103/PhysRevD.81.054028}{\emph{Phys.
  Rev.} {\bf D81} (2010) 054028}, [\href{http://arxiv.org/abs/1001.5034}{{\tt
  1001.5034}}].

\bibitem{Aliev:2010zk}
M.~Aliev, H.~Lacker, U.~Langenfeld, S.~Moch, P.~Uwer and M.~Wiedermann,
  \emph{{HATHOR: HAdronic Top and Heavy quarks crOss section calculatoR}},
  \href{http://dx.doi.org/10.1016/j.cpc.2010.12.040}{\emph{Comput. Phys.
  Commun.} {\bf 182} (2011) 1034--1046},
  [\href{http://arxiv.org/abs/1007.1327}{{\tt 1007.1327}}].

\bibitem{Kant:2014oha}
P.~Kant, O.~M. Kind, T.~Kintscher, T.~Lohse, T.~Martini, S.~M{\"o}lbitz et~al.,
  \emph{{HatHor for single top-quark production: Updated predictions and
  uncertainty estimates for single top-quark production in hadronic
  collisions}},
  \href{http://dx.doi.org/10.1016/j.cpc.2015.02.001}{\emph{Comput. Phys.
  Commun.} {\bf 191} (2015) 74--89},
  [\href{http://arxiv.org/abs/1406.4403}{{\tt 1406.4403}}].

\bibitem{Alekhin:2013nda}
S.~Alekhin, J.~Blumlein and S.~Moch, \emph{{The ABM parton distributions tuned
  to LHC data}},
  \href{http://dx.doi.org/10.1103/PhysRevD.89.054028}{\emph{Phys. Rev.} {\bf
  D89} (2014) 054028}, [\href{http://arxiv.org/abs/1310.3059}{{\tt
  1310.3059}}].

\bibitem{CDF:2014uma}
{\scshape CDF, D0} collaboration, T.~A. Aaltonen et~al., \emph{{Observation of
  s-channel production of single top quarks at the Tevatron}},
  \href{http://dx.doi.org/10.1103/PhysRevLett.112.231803}{\emph{Phys. Rev.
  Lett.} {\bf 112} (2014) 231803}, [\href{http://arxiv.org/abs/1402.5126}{{\tt
  1402.5126}}].

\bibitem{Melnikov:2000qh}
K.~Melnikov and T.~v. Ritbergen, \emph{{The Three loop relation between the
  MS-bar and the pole quark masses}},
  \href{http://dx.doi.org/10.1016/S0370-2693(00)00507-4}{\emph{Phys. Lett.}
  {\bf B482} (2000) 99--108}, [\href{http://arxiv.org/abs/hep-ph/9912391}{{\tt
  hep-ph/9912391}}].

\end{thebibliography}\endgroup
\bibliographystyle{JHEP}

\end{document}